\documentclass{article}
\usepackage{smc2026}
\usepackage[T1]{fontenc}
\usepackage[utf8]{inputenc}
\usepackage{amsmath,cite,url}
\usepackage{graphicx}
\usepackage{color}
\usepackage{booktabs}
\usepackage[caption=false, font=footnotesize]{subfig}
\usepackage{paralist}
\usepackage[figure,table]{hypcap}
\usepackage{multirow}

\def\papertitle{BMdataset: A Musicologically Curated LilyPond Dataset}

\author[1]{\mbox{\firstname{Matteo}\lastname{Spanio}\email{spanio@dei.unipd.it}}}
\author[2]{\mbox{\firstname{Ilay}\lastname{Guler}\email{ilay2004@bu.edu}}}
\author[1]{\mbox{\firstname{Antonio}\lastname{Rod\'a}\email{roda@dei.unipd.it}}}

\affil[1]{\department{Department of Information Engineering}\institution{University of Padua}\city{Padua}\country{Italy}\affiliationtype{University}}
\affil[2]{\institution{Boston University}\city{Boston}\country{USA}\affiliationtype{University}}

\completesetup
\title{\papertitle}

\begin{document}
\capstartfalse
\maketitle
\capstarttrue

\begin{abstract}
Symbolic music research has relied almost exclusively on MIDI-based datasets; text-based engraving formats such as LilyPond remain unexplored for music understanding. We present \textbf{BMdataset}, a musicologically curated dataset of 347 LilyPond scores (2,646 movements) transcribed by experts directly from original Baroque manuscripts, with metadata covering composer, musical form, instrumentation, and sectional attributes. Building on this resource, we introduce \textbf{LilyBERT}\footnote{\url{https://huggingface.co/csc-unipd/lilybert}}, a CodeBERT-based encoder adapted to symbolic music through vocabulary extension with 115 LilyPond-specific tokens and masked language model pre-training. Linear probing on the out-of-domain Mutopia corpus shows that, despite its modest size (${\sim}$90M tokens), fine-tuning on BMdataset alone outperforms continuous pre-training on the full PDMX corpus (${\sim}$15B tokens) for both composer and style classification, demonstrating that small, expertly curated datasets can be more effective than large, noisy corpora for music understanding. Combining broad pre-training with domain-specific fine-tuning yields the best results overall (84.3\% composer accuracy), confirming that the two data regimes are complementary. We release the dataset, tokenizer, and model to establish a baseline for representation learning on LilyPond.
\end{abstract}

\section{Introduction}\label{sec:introduction}

Recent work in music information retrieval (MIR) and generative AI has renewed attention to symbolic music representations~\cite{dong2023mmt, zeng2021musicbert, thickstun2023anticipatory}. While audio-domain models have seen rapid commercial development, symbolic formats remain essential for musicological analysis, score generation, and artist-in-the-loop workflows where fine-grained control over musical structure is required.

Among symbolic formats, MIDI dominates: large-scale datasets such as Lakh MIDI~\cite{raffel2016learning} and PDMX~\cite{long2024pdmx} support classification, generation, and retrieval tasks. However, MIDI discards much of the notational information present in written scores, including engraving directives, section boundaries, and the hierarchical structure of multi-movement works. MusicXML partially addresses this gap, but no comparable resources exist for \emph{LilyPond}\footnote{\url{https://lilypond.org}}, a widely used open-source music engraving system whose text-based markup language encodes both musical content and typographic structure in a human-readable, version-controllable format.

LilyPond's syntax shares structural properties with programming languages: it uses nested blocks, backslash commands, variable assignments, and macro definitions. This observation motivates the use of \emph{code pre-trained} language models, which have been shown to capture hierarchical and syntactic patterns in structured text~\cite{feng2020codebert}, as a foundation for learning LilyPond representations.

In this paper, we make four contributions:
\begin{enumerate}
    \item \textbf{BMdataset}, a musicologically curated dataset of 347 compilable LilyPond scores (2,646 movements) sourced from BaroqueMusic.it\footnote{\url{https://www.baroquemusic.it}, last visited: november 2025}, transcribed directly from original manuscripts. The dataset is accompanied by structured metadata (composer, musical form, instrumentation, key, tempo, time signature, and historical period) and will be released with \texttt{.ly}, \texttt{.midi}, \texttt{.pdf}, and \texttt{.json} metadata files.
    \item A \textbf{LilyPond tokenizer} that extends the CodeBERT vocabulary with 115 domain-specific tokens, ensuring that musically meaningful commands (e.g., \texttt{\textbackslash relative}, \texttt{\textbackslash clef}, \texttt{\textbackslash key}) are represented as atomic units rather than fragmented subwords.
    \item \textbf{LilyBERT}, a CodeBERT-based encoder adapted to symbolic music through masked language model (MLM) pre-training on LilyPond corpora.
    \item A systematic \textbf{probing evaluation} on the Mutopia corpus comparing four model variants that isolate the effects of pre-training data source, corpus size, and domain-specific fine-tuning on downstream composer and style classification.
\end{enumerate}

\section{Related Work}\label{sec:related}

\subsection{Symbolic Music Datasets}

\begin{table}[t]
\centering
\caption{Comparison of symbolic music datasets. BMdataset is the first musicologically curated LilyPond-native dataset.}
\footnotesize
\begin{tabular}{llrl}
    \toprule
    \textbf{Dataset} & \textbf{Format} & \textbf{Size} & \textbf{License} \\
    \midrule
    LMD~\cite{raffel2016learning} & MIDI & 174K & CC-BY$^\dagger$ \\
    SymphonyNet~\cite{liu2022symphony} & MIDI & 46K & -- \\
    MAESTRO~\cite{hawthorne2018enabling} & MIDI & 1.3K & CC-BY-NC-SA \\
    Wikifonia~\cite{wikifonia} & MusicXML & 6.4K & -- \\
    POP909~\cite{wang2020pop909} & MIDI & 909 & -- \\
    Nottingham~\cite{boulanger2012modeling} & ABC & 1.2K & PD \\
    ABC-Tune~\cite{wu2023tunesformer} & ABC & 165K & CC-0 \\
    PDMX~\cite{long2024pdmx} & MusicXML & 254K & CC-0 \\
    Mutopia & LilyPond & 2.1K & PD \\
    \midrule
    \textbf{BMdataset} & \textbf{LilyPond} & \textbf{347} & \textbf{CC-BY-NC-ND} \\
    \bottomrule
    \multicolumn{4}{l}{\footnotesize $\dagger$ Contains documented copyrighted works~\cite{thickstun2023anticipatory}.}
\end{tabular}
\label{tab:datasets}
\end{table}

Table~\ref{tab:datasets} summarises existing symbolic music datasets. Most large-scale datasets use MIDI~\cite{raffel2016learning, liu2022symphony, hawthorne2018enabling, wang2020pop909} or MusicXML~\cite{long2024pdmx, wikifonia} formats. MIDI is compact and widely supported but discards notational details such as engraving directives, articulation placement, and the hierarchical structure of multi-movement works. MusicXML preserves richer score information but remains tied to XML-based tooling. LilyPond, by contrast, represents music as structured plaintext --- making it version-controllable, human-readable, and directly processable by language models without format conversion. While the Mutopia Project provides a community-curated collection of public domain LilyPond scores, it lacks structured metadata beyond basic composer and style tags and has not been used as a basis for representation learning. To our knowledge, no musicologically curated LilyPond dataset with systematic metadata annotation currently exists.

\subsection{Music Representation Learning}

Pre-trained Transformer models for symbolic music have improved both classification and generation benchmarks. MusicBERT~\cite{zeng2021musicbert} and MidiBERT-Piano~\cite{chou2021midibert} adapt BERT-style pre-training to MIDI representations and report strong downstream classification performance. Tokenisation-focused work such as Byte Pair Encoding for Symbolic Music~\cite{fradet2023bpe} demonstrates that compression-aware vocabularies can improve both quality and efficiency in MIDI-based modelling. On the generative side, MIDI-GPT~\cite{pasquier2025midigpt} shows controllable multitrack composition in MIDI, while NotaGen~\cite{wang2025notagen} scales symbolic generation with ABC notation pre-training.

Most of these pipelines operate on MIDI. The principal text-based alternative, ABC notation\footnote{\url{https://abcnotation.com}}, is more limited in structural and engraving expressivity, and parts of the ABC 2.1 specification are still marked as beta~\cite{abcstandard21}. LilyPond provides a richer score representation (layout directives, reusable variables/macros, and hierarchical score structure) that has not yet been used for pre-training.

\subsection{Code Pre-trained Models}

CodeBERT~\cite{feng2020codebert} is a bimodal pre-trained model for programming languages and natural language, based on the RoBERTa architecture~\cite{liu2019roberta}. Pre-trained on six programming languages using both MLM and replaced-token-detection (RTD) objectives, CodeBERT achieves strong results on code search, documentation generation, and clone detection. Subsequent work has incorporated richer structural signals: GraphCodeBERT~\cite{guo2021graphcodebert} augments the input with data-flow graphs, while AST-aware approaches encode parse-tree structure directly. Because LilyPond files are compiled by a strict grammar, they could expose parse-tree or block-hierarchy information to such structure-aware encoders in future work.

Our motivation for selecting CodeBERT as the backbone is that LilyPond behaves more like a compiled formal language than a simple markup format: files are parsed according to a strict grammar and compiled into engraved score outputs. Its syntax employs code-like constructs such as nested blocks (\texttt{\{ \}}), backslash commands (\texttt{\textbackslash relative}), variable bindings, includes, and macro definitions. We therefore hypothesise that a code-pretrained encoder is a better inductive fit than a model trained only on plain text token statistics. This parallels recent work showing that code-trained models can be effective on non-code structured languages such as chemical notation and mathematical expressions~\cite{dagan2024tokenizer}.

\section{BMdataset Dataset}\label{sec:dataset}

\subsection{Source and Compilation}

The BMdataset dataset originates from BaroqueMusic.it, a collection of 383 LilyPond projects transcribed by musicologists who worked directly from original manuscript sources. Each transcription is annotated with a reference to the original manuscript and its catalogue number, so that every score can be traced to its primary source. The majority of the collection consists of works by lesser-known Baroque composers; to our knowledge, this is the first musicologically curated LilyPond dataset focused on under-represented figures of the period.

Each project is organised as a multi-file LilyPond workspace comprising: (i) a macro-definition file with shorthand musical markups; (ii) a header file specifying movement order via \texttt{\textbackslash include} directives; (iii) individual movement files containing instrument-specific variables within \texttt{\textbackslash relative} blocks; and (iv) a score file orchestrating all instruments across movements.

To produce self-contained files suitable for modelling, we parsed each project's header to extract the dependency order, concatenated the movement files sequentially, and appended the score file. The resulting files were validated by compiling each one with the LilyPond executable and scanning \texttt{stderr} for error messages. Of the original 383 projects, 36 failed compilation due to syntax errors or unsupported characters, yielding \textbf{347 compilable LilyPond files} comprising \textbf{2,646 individual movements} (i.e., self-contained musical units such as \emph{Allegro}, \emph{Adagio}), which in turn contain \textbf{1,679 tempo-marked sections} (passages delimited by tempo or character indications within a movement). The corpus totals approximately 90M tokens as measured by our extended tokenizer.

\subsection{Metadata and Labels}\label{sec:labels}

Each file is annotated with the following metadata, stored in a structured JSON manifest:

\begin{itemize}
    \item \textbf{Composer} (71 unique): extracted from filename conventions and standardised.
    \item \textbf{Musical form} (16 unique): matched against a predefined list of Baroque forms (concerto, sonata, opera, suite, aria, cantata, etc.).
    \item \textbf{MIDI instruments} (25 unique): extracted via regex from \texttt{\textbackslash set Staff.midiInstrument} directives.
    \item \textbf{Section metadata}: for each section (movement, dance, etc.), we record key, scale (major/minor), tempo, and time signature, extracted from \texttt{forma} variable blocks.
    \item \textbf{Historical period}: Early Baroque ($<$1650), High Baroque (1650--1700), Late Baroque (1700--1750), and Transitional Classical ($>$1750).
\end{itemize}

Section names were classified into semantically meaningful categories using a curated taxonomy: \emph{speed} labels (e.g., allegro, largo), \emph{intention} labels (e.g., affettuoso, maestoso), \emph{suite} dances (e.g., bourr\'{e}e, sarabande), \emph{no-tempo} labels (e.g., recitativo, ouverture), and \emph{non-descriptive} labels (e.g., aria, trio). For sections with uninformative names, quarter-note BPM was computed and mapped to standard tempo categories. This taxonomy is formalised as a directed acyclic graph (DAG) to preserve overlapping semantic relationships --- for instance, suite dances such as \emph{giga} belong to both the \emph{suite} and \emph{fast} speed categories.

\subsection{Dataset Statistics}

\begin{figure*}[t]
\centering
\subfloat[Musical form distribution]{
    \includegraphics[width=0.32\textwidth]{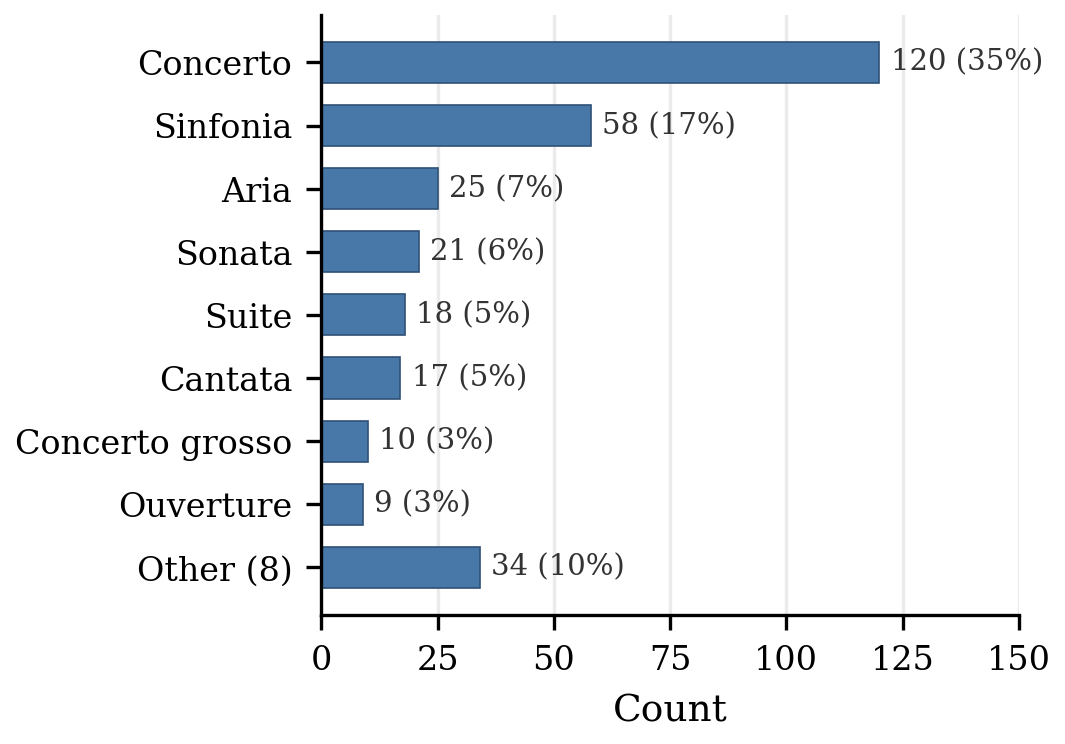}
}
\hfill
\subfloat[Musical period distribution]{
    \includegraphics[width=0.32\textwidth]{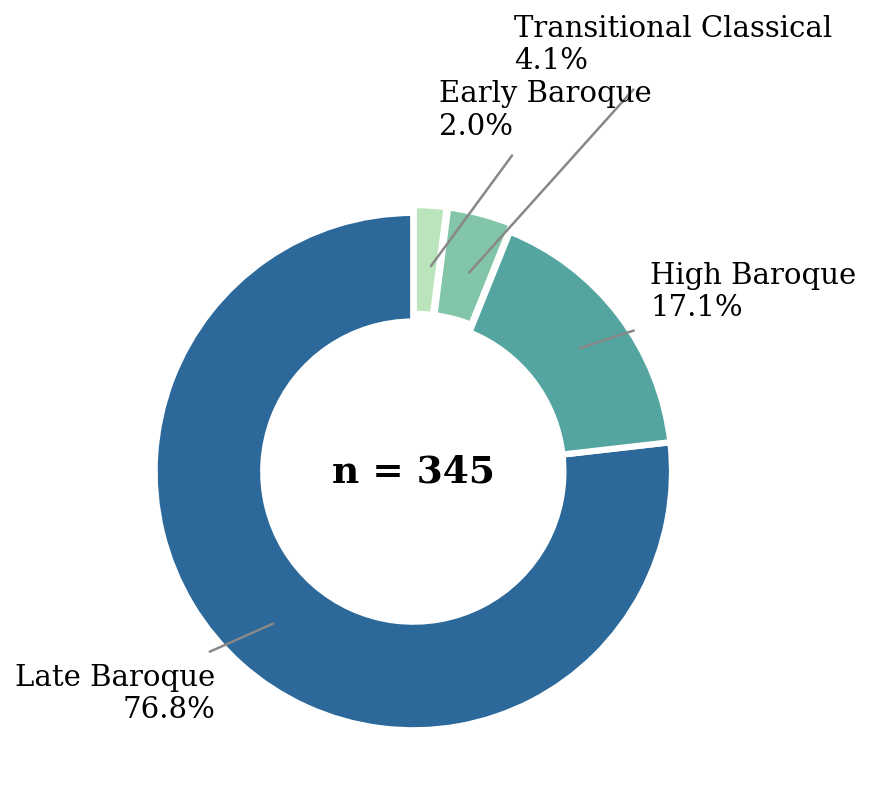}
}
\hfill
\subfloat[Instrument distribution]{
    \includegraphics[width=0.32\textwidth]{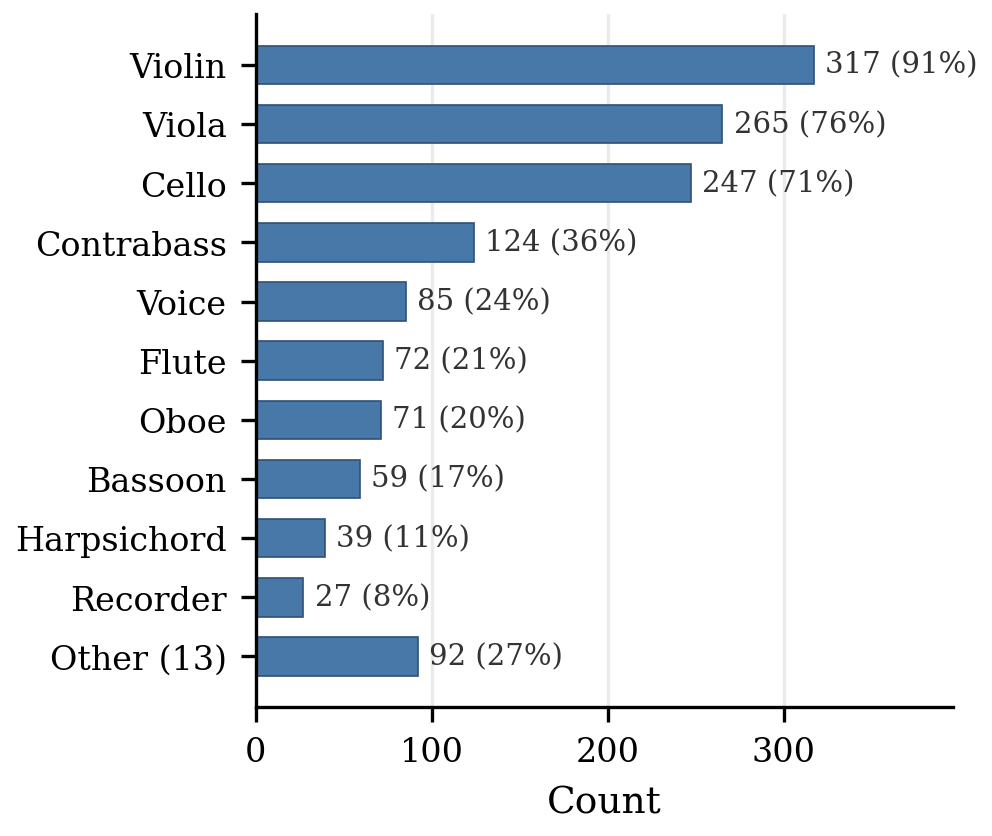}
}

\vspace{0.5em}

\subfloat[Key distribution]{
    \includegraphics[width=0.32\textwidth]{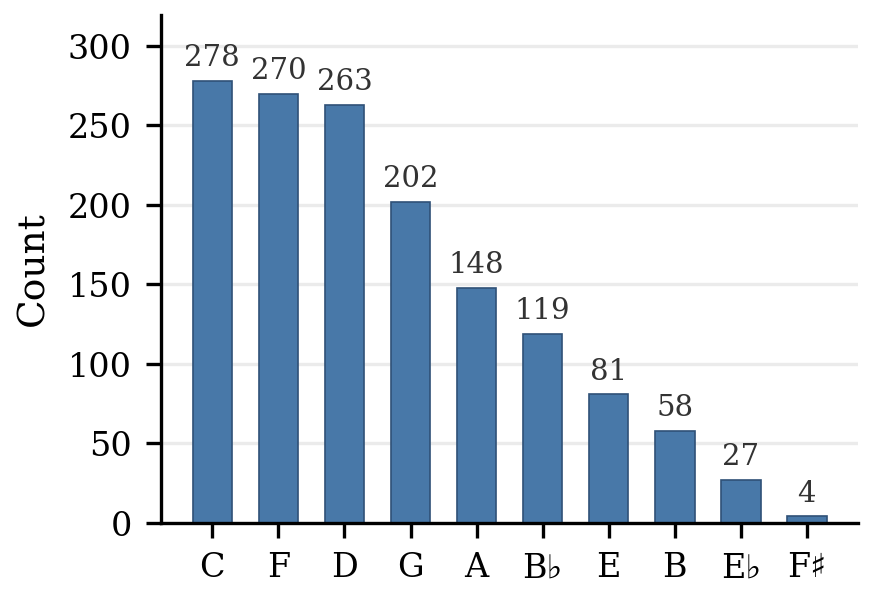}
}
\hfill
\subfloat[Time signature distribution]{
    \includegraphics[width=0.32\textwidth]{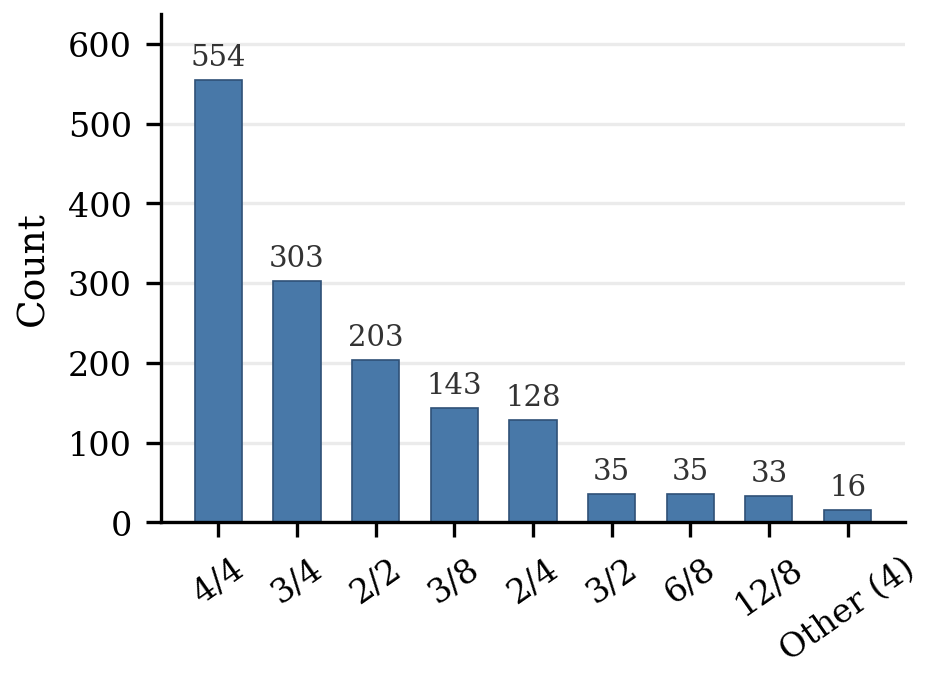}
}
\hfill
\subfloat[Composer distribution (top 10)]{
    \includegraphics[width=0.32\textwidth]{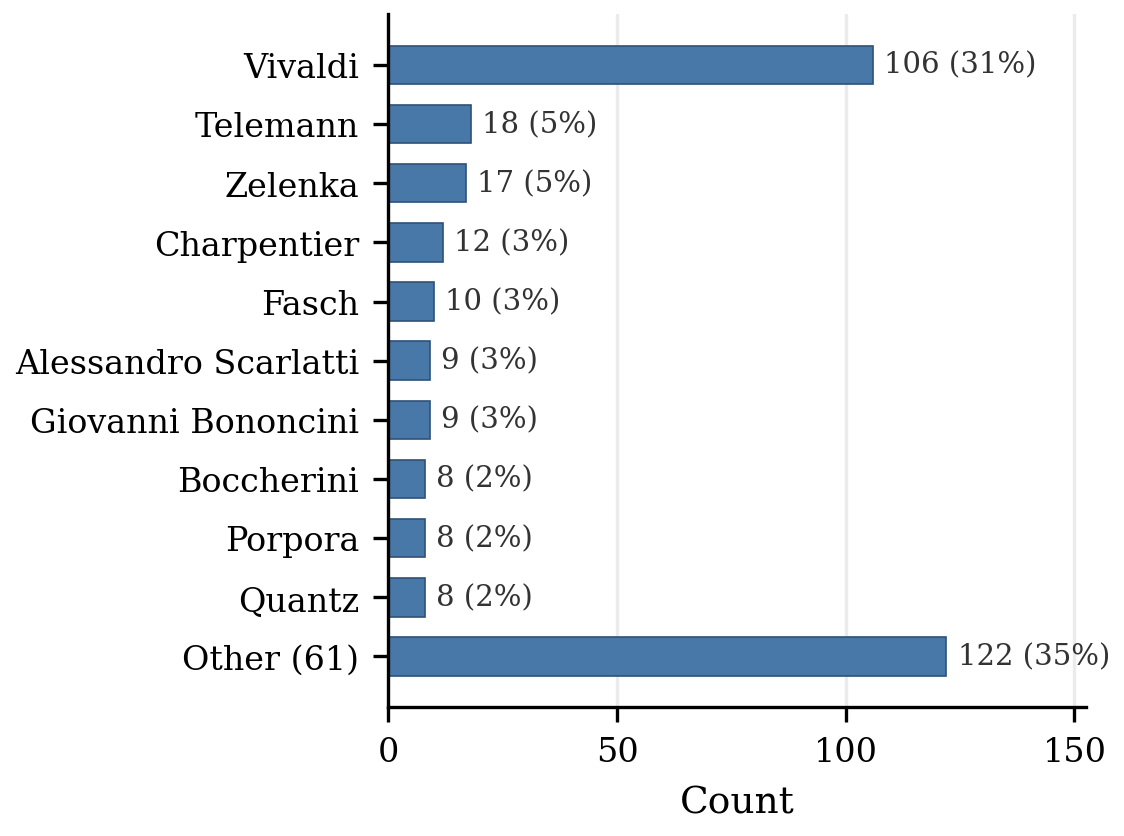}
}
\caption{BMdataset dataset statistics. The dataset is dominated by Late Baroque works, string-centric instrumentation, and concerto forms. Vivaldi accounts for 30.5\% of compositions, while the remaining 70 composers contribute 1--18 pieces each.}
\label{fig:dataset_statistics}
\end{figure*}

Figure~\ref{fig:dataset_statistics} summarises the dataset distributions. The dataset exhibits characteristic imbalances: the composer distribution is heavily skewed towards Vivaldi (106 pieces, 30.5\%), string instruments dominate the instrumentation (violin appears in 91.6\% of files), and Late Baroque works (1700--1750) constitute 76.4\% of the corpus. Concertos are the most common form (34.6\%), followed by sinfonias (16.7\%). The 1,679 individual sections span 65 unique section names, with speed-related labels being most frequent (1,212 occurrences across slow, mid, fast, and very fast categories), followed by suite dances (232) and non-tempo labels (137). The dataset nonetheless spans 71 composers, 16 musical forms, 25 instruments, 10 keys, and 12 time signatures.

What sets BMdataset apart from crowd-sourced alternatives is provenance. Each transcription was produced from the original manuscript, and the metadata records the manuscript reference and catalogue number for every score. This chain of evidence ties the musical content to the historical source, rather than to the editorial interpretations or modern reharmonisations found in community-contributed corpora.

\subsection{Parser Validation}

To verify the accuracy of the parsed musical content, we developed a batch-testing pipeline that compares the parser's note count against the number of rendered notehead glyphs in compiled PostScript outputs. Each \texttt{.ly} file was compiled with the \texttt{--ps} flag, and notehead glyphs were extracted via a regex search for \texttt{/noteheads.} patterns. After accounting for unused variable blocks (e.g., cadenza variables defined but never called in the score) and editorial incipits, 332 out of 347 files (95.7\%) achieved a perfect 1:1 match. The remaining 15 files exhibited a mean error of 0.26\%, attributable to rare LilyPond formatting edge cases.

\section{LilyBERT Model and Tokenizer}\label{sec:model}

\subsection{Tokenizer Design}\label{sec:tokenizer}

LilyPond's syntax relies heavily on backslash-prefixed commands (e.g., \texttt{\textbackslash relative}, \texttt{\textbackslash clef}, \texttt{\textbackslash key}, \texttt{\textbackslash time}), which carry strong musical semantics. A standard BPE tokenizer, as used in CodeBERT, fragments these commands into meaningless subwords. For example, \texttt{\textbackslash relative} might be split into \texttt{\textbackslash}, \texttt{rel}, \texttt{ative} --- destroying the semantic unit.

To address this, we extend the CodeBERT tokenizer (RoBERTa BPE, 50,265 vocabulary) with 115 LilyPond-specific tokens using the HuggingFace \texttt{add\_tokens()} mechanism. The added tokens fall into six categories:

\begin{itemize}
    \item \textbf{Musical commands} (15): \texttt{\textbackslash time}, \texttt{\textbackslash key}, \texttt{\textbackslash clef}, \texttt{\textbackslash tempo}, \texttt{\textbackslash repeat}, \texttt{\textbackslash tuplet}, \texttt{\textbackslash grace}, \texttt{\textbackslash partial}, etc.
    \item \textbf{Dynamics} (19): \texttt{\textbackslash p}, \texttt{\textbackslash pp}, \texttt{\textbackslash f}, \texttt{\textbackslash ff}, \texttt{\textbackslash mp}, \texttt{\textbackslash mf}, \texttt{\textbackslash sfz}, \texttt{\textbackslash cresc}, \texttt{\textbackslash decresc}, etc.
    \item \textbf{Structural blocks} (20): \texttt{\textbackslash score}, \texttt{\textbackslash relative}, \texttt{\textbackslash absolute}, \texttt{\textbackslash transpose}, \texttt{\textbackslash header}, \texttt{\textbackslash layout}, \texttt{\textbackslash midi}, \texttt{\textbackslash markup}, etc.
    \item \textbf{Articulations/ornaments} (14): \texttt{\textbackslash trill}, \texttt{\textbackslash fermata}, \texttt{\textbackslash mordent}, \texttt{\textbackslash staccato}, \texttt{\textbackslash accent}, \texttt{\textbackslash arpeggio}, etc.
    \item \textbf{Key modes} (9): \texttt{\textbackslash major}, \texttt{\textbackslash minor}, \texttt{\textbackslash dorian}, \texttt{\textbackslash phrygian}, \texttt{\textbackslash lydian}, \texttt{\textbackslash mixolydian}, etc.
    \item \textbf{Property overrides and other} (38): \texttt{\textbackslash override}, \texttt{\textbackslash set}, \texttt{\textbackslash once}, \texttt{\textbackslash segno}, \texttt{\textbackslash coda}, pedal marks, voice directives, etc.
\end{itemize}

The resulting tokenizer has a vocabulary of 50,380 tokens. Each added token is treated as a single, unsplittable unit during tokenisation, so musically meaningful commands remain atomic. Although the underlying RoBERTa architecture natively uses \texttt{<s>} and \texttt{</s>} as special tokens, our extended tokenizer remaps these to \texttt{[CLS]} and \texttt{[SEP]} for consistency with the BERT convention used in the probing literature.

\subsection{Model Architecture}

LilyBERT is based on CodeBERT (\texttt{microsoft/codebert-base})~\cite{feng2020codebert}, a RoBERTa-based~\cite{liu2019roberta} encoder with 12 transformer layers, 768 hidden dimensions, 12 attention heads, and approximately 125M parameters. We resize the token embedding matrix from 50,265 to 50,380 to accommodate the added LilyPond vocabulary. The new embedding rows are randomly initialised and learned during pre-training.

As discussed in Section~\ref{sec:related}, LilyPond shares structural properties with programming languages (nested blocks, command--argument patterns, variable definitions, macro expansion), which makes CodeBERT's pre-training on six programming languages a natural starting point.

\subsection{Pre-training Procedure}

LilyBERT is trained using the standard masked language modelling (MLM) objective: 15\% of input tokens are randomly masked, and the model is trained to predict the original token from context. Although CodeBERT's original pre-training additionally employs a replaced-token-detection (RTD) objective, RTD requires a separate generator network and has been shown to provide marginal gains over MLM alone in CodeBERT's own ablations~\cite{feng2020codebert}. We therefore use MLM only, which simplifies the training pipeline and avoids the need to train an auxiliary generator on our domain-specific vocabulary.

To ensure consistent pitch notation across all corpora, the BMdataset scores (originally in Italian notation) are converted to LilyPond's \texttt{nederlands} pitch language prior to training. This matches the converted PDMX scores, where \texttt{musicxml2ly} defaults to \texttt{nederlands}, as well as the majority of files in the Mutopia evaluation corpus.

Training proceeds in two stages:
\begin{enumerate}
    \item \textbf{Continuous pre-training on PDMX-LilyPond}: We convert a large subset of the PDMX dataset~\cite{long2024pdmx} from MusicXML to LilyPond format using \texttt{musicxml2ly}, yielding approximately 15 billion tokens. The model is trained for 1 epoch on this corpus.
    \item \textbf{Fine-tuning on BMdataset}: The model is further fine-tuned on the BMdataset corpus (${\sim}$90M tokens) for 10 epochs.
\end{enumerate}

Training uses the following hyperparameters: effective batch size of 288 across 4 NVIDIA A40 GPUs (48\,GB) with Distributed Data Parallel (DDP), learning rate $2 \times 10^{-4}$ with cosine schedule and 10\% warmup, weight decay 0.01, gradient clipping at 1.0, \texttt{bfloat16} mixed precision, AdamW optimiser, and a maximum sequence length of 512 tokens. Early stopping with patience 5 is applied based on validation loss. Input files are split into non-overlapping chunks of 512 tokens, each wrapped with \texttt{[CLS]} and \texttt{[SEP]} tokens.

\section{Experiments}\label{sec:experiments}

\begin{table*}[ht!]
\centering
\caption{Linear probing results (layer 6, mean $\pm$ std over 5 folds). Best in \textbf{bold}, second best \underline{underlined}.}
\footnotesize
\begin{tabular}{l ccc ccc}
    \toprule
    & \multicolumn{3}{c}{\textbf{Composer (30 classes)}} & \multicolumn{3}{c}{\textbf{Style (10 classes)}} \\
    \cmidrule(lr){2-4} \cmidrule(lr){5-7}
    \textbf{Model} & \textbf{Acc.} & \textbf{Prec.} & \textbf{Recall} & \textbf{Acc.} & \textbf{Prec.} & \textbf{Recall} \\
    \midrule
    CB + PDMX\textsubscript{full}          & $.808{\scriptstyle\pm.021}$ & $.751{\scriptstyle\pm.025}$ & $.751{\scriptstyle\pm.032}$ & $.826{\scriptstyle\pm.015}$ & $\mathbf{.838}{\scriptstyle\pm.044}$ & $.781{\scriptstyle\pm.031}$ \\
    CB + BMdataset                       & $\underline{.829}{\scriptstyle\pm.013}$ & $.750{\scriptstyle\pm.039}$ & $\underline{.752}{\scriptstyle\pm.036}$ & $\mathbf{.837}{\scriptstyle\pm.004}$ & $.813{\scriptstyle\pm.047}$ & $\underline{.794}{\scriptstyle\pm.016}$ \\
    CB + PDMX\textsubscript{90M}           & $.817{\scriptstyle\pm.017}$ & $\mathbf{.767}{\scriptstyle\pm.035}$ & $.747{\scriptstyle\pm.036}$ & $.823{\scriptstyle\pm.009}$ & $.805{\scriptstyle\pm.053}$ & $.775{\scriptstyle\pm.039}$ \\
    CB + PDMX $\rightarrow$ BM             & $\mathbf{.843}{\scriptstyle\pm.010}$ & $\underline{.765}{\scriptstyle\pm.031}$ & $\mathbf{.775}{\scriptstyle\pm.028}$ & $\underline{.829}{\scriptstyle\pm.011}$ & $\underline{.825}{\scriptstyle\pm.013}$ & $\mathbf{.797}{\scriptstyle\pm.035}$ \\
    \bottomrule
\end{tabular}
\label{tab:results}
\end{table*}

\subsection{Probing Setup}

To evaluate whether LilyBERT's representations encode musically meaningful properties, we conduct linear probing experiments on the Mutopia corpus --- a community-maintained collection of 2,123 public-domain LilyPond files spanning 320 composers and 13 musical styles. Mutopia is independent from BMdataset and covers a broader range of periods and styles, providing an out-of-domain evaluation setting.

\textbf{Tasks.} We evaluate two classification tasks: (1) \textbf{composer classification} and (2) \textbf{style/period classification}. To ensure sufficient training examples, we filter out classes with fewer than 10 pieces, yielding 30 composer classes (1,438 samples) and 10 style classes (1,972 samples).

\textbf{Probe architecture.} To prevent the model from exploiting authorship metadata, LilyPond file headers (containing composer name and revision history) are stripped prior to embedding extraction. Each file is split into non-overlapping 512-token chunks and processed by a frozen model; token-level hidden states are mean-pooled across all chunks to produce a single embedding per file at layers 3, 6, 9, and 12. Embeddings are standardised with \texttt{StandardScaler} and fed to a linear classifier (MLP with no hidden layers). We report results using 5-fold stratified cross-validation.

\textbf{Metrics.} We report accuracy, and macro-averaged precision and recall.

\subsection{Models Compared}

To isolate the effects of pre-training data source, corpus size, and domain-specific fine-tuning, we compare four model variants:

\begin{enumerate}
    \item \textbf{CodeBERT + PDMX\textsubscript{full}}: CodeBERT with continuous MLM pre-training on the full PDMX-LilyPond corpus (${\sim}$15B tokens, 1 epoch).
    \item \textbf{CodeBERT + BMdataset}: CodeBERT fine-tuned on BMdataset (${\sim}$90M tokens, 10 epochs).
    \item \textbf{CodeBERT + PDMX\textsubscript{90M}}: CodeBERT fine-tuned on a size-matched PDMX-LilyPond subset (${\sim}$90M tokens, 10 epochs), enabling a fair comparison with variant 2 that controls for corpus size.
    \item \textbf{CodeBERT + PDMX\textsubscript{full} $\rightarrow$ BMdataset}: two-stage training --- continuous pre-training on full PDMX followed by fine-tuning on BMdataset (10 epochs).
\end{enumerate}

Comparing variants 2 and 3 directly tests whether the domain specificity of BMdataset (musicologically curated, Baroque-focused) provides an advantage over a generic LilyPond corpus of equal size. Variant 4 tests whether large-scale pre-training followed by domain-specific fine-tuning is complementary.

\subsection{Results}

\begin{figure}[t]
\centering
\includegraphics[width=\columnwidth]{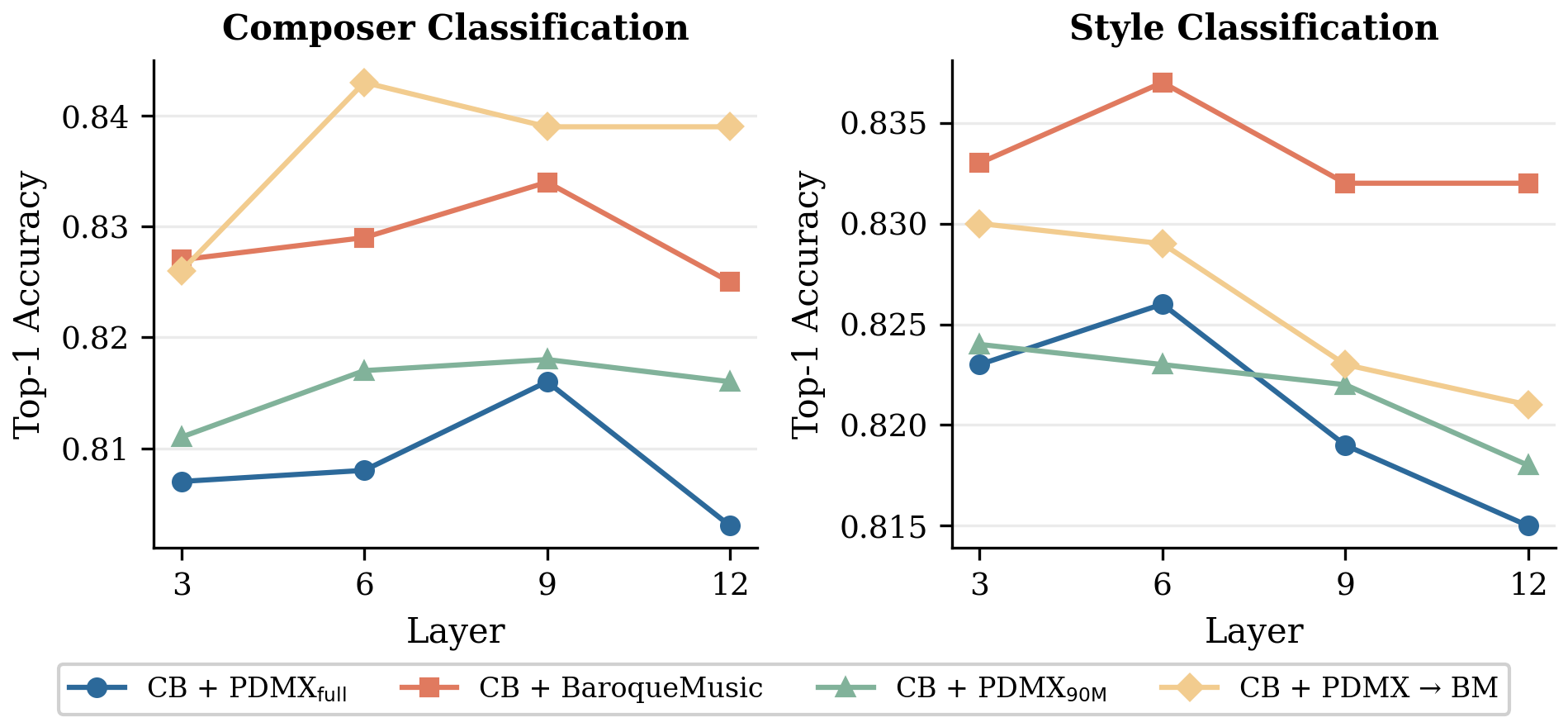}
\caption{Layer-wise probing accuracy for composer and style classification. Composer information is best captured at middle layers (6--9), while style features are strongest in earlier layers and degrade towards the output.}
\label{fig:layer_accuracy}
\end{figure}

\begin{figure}[t]
\centering
\includegraphics[width=\columnwidth]{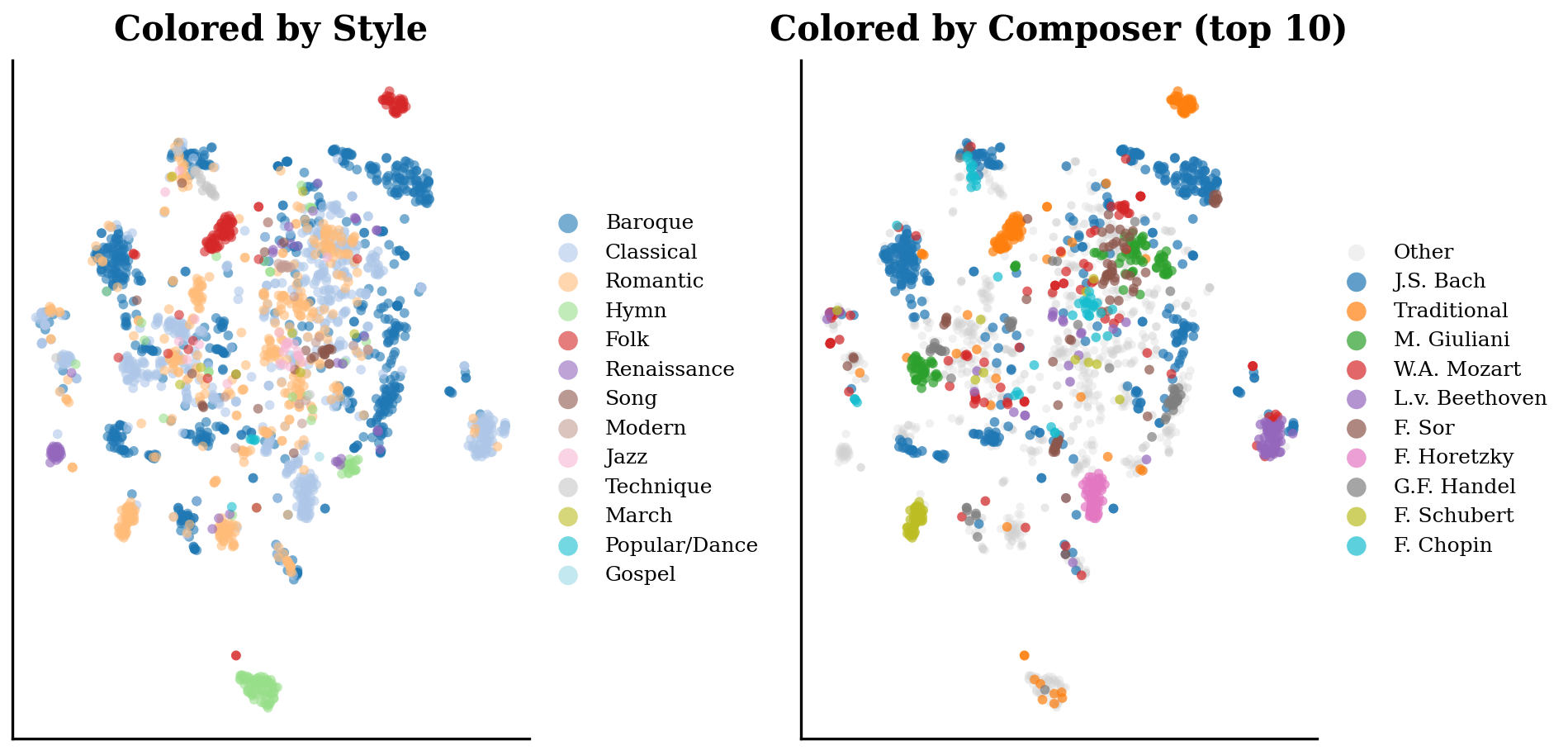}
\caption{t-SNE projection of layer-6 embeddings (CB + PDMX $\rightarrow$ BM) on Mutopia, colored by style (left) and composer (right). Stylistic clusters (Baroque, Classical, Romantic) are clearly separated; within-style composer clusters indicate that the model captures both global period conventions and individual compositional idioms.}
\label{fig:tsne}
\end{figure}

\begin{figure}[t]
\centering
\includegraphics[width=\columnwidth]{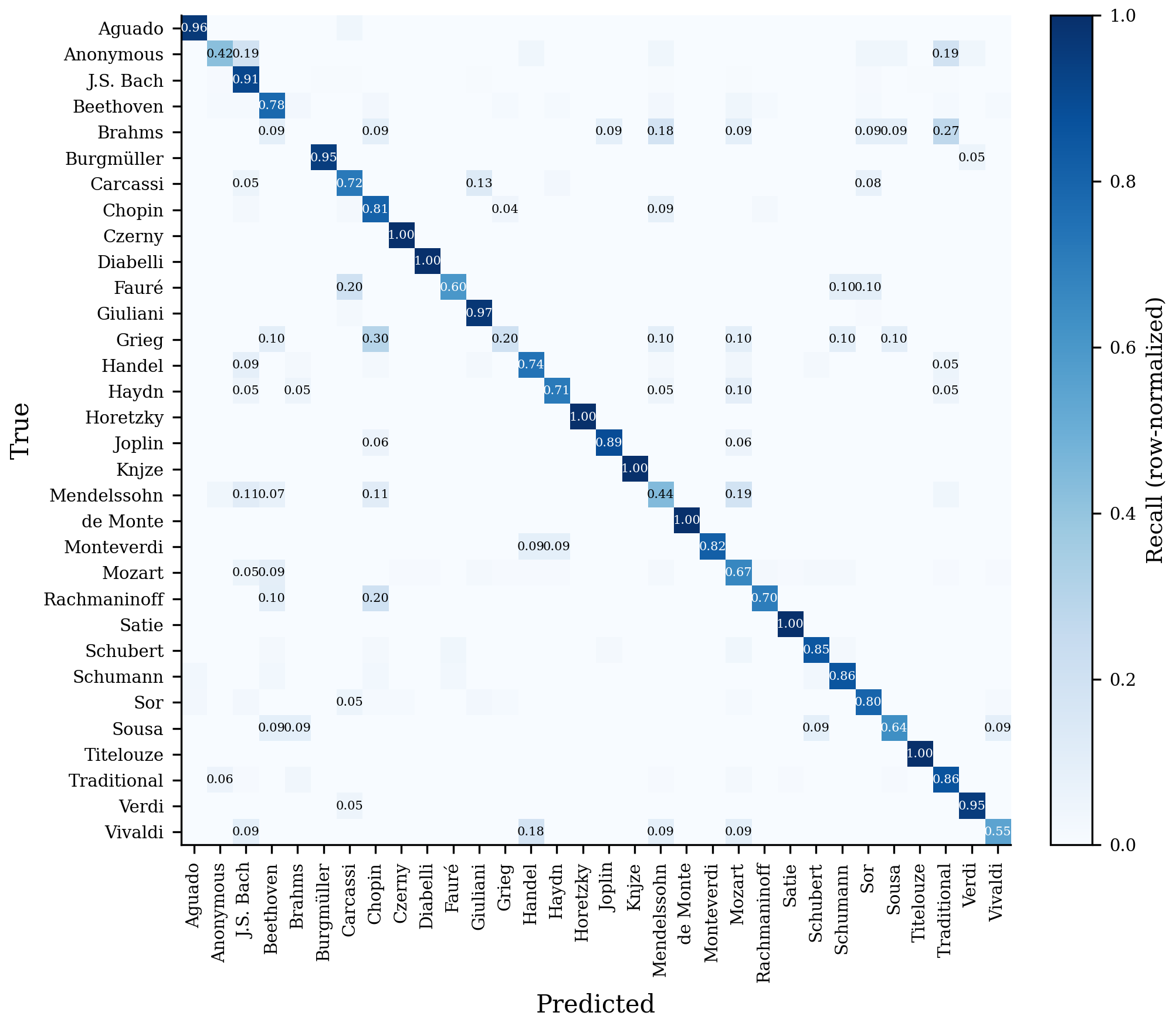}
\caption{Row-normalised confusion matrix for composer classification (CB + PDMX $\rightarrow$ BM, layer 6, aggregated over 5 folds). Most misclassifications occur between stylistically related composers (e.g., Grieg $\leftrightarrow$ Chopin, Haydn $\leftrightarrow$ Mozart), suggesting that errors reflect genuine stylistic proximity rather than random noise.}
\label{fig:confusion}
\end{figure}

Table~\ref{tab:results} presents the probing results for composer and style classification at layer 6, while Figure~\ref{fig:layer_accuracy} shows performance across all probed layers. Figure~\ref{fig:tsne} visualises the learned embedding space, and Figure~\ref{fig:confusion} shows the composer confusion matrix for the best model.

\textbf{Domain specificity vs.\ scale.}
Comparing CB~+~BMdataset against CB~+~PDMX\textsubscript{90M} --- two models trained on corpora of equal size (${\sim}$90M tokens) --- reveals a consistent advantage for the musicologically curated data: BMdataset yields higher accuracy on both composer (+1.2pp) and style (+1.4pp) classification. This suggests that expert curation and stylistic coherence in the training corpus are more valuable than raw diversity when the downstream task requires fine-grained musical discrimination.

\textbf{Complementarity of pre-training stages.}
The two-stage pipeline CB~+~PDMX~$\rightarrow$~BM achieves the highest composer accuracy (0.843) and the best recall on both tasks, outperforming either pre-training stage alone. This supports a curriculum learning interpretation: broad-coverage pre-training on PDMX establishes general LilyPond literacy, while domain-specific fine-tuning on BMdataset sharpens the model's sensitivity to stylistic features. For style classification, however, the gains are more modest, with CB~+~BMdataset remaining the best model by top-1 accuracy (0.837). This asymmetry may reflect the fact that composer identity depends on fine-grained idiosyncratic patterns that benefit from large-scale pre-training, whereas broader stylistic categories (e.g., Baroque, Classical) are already well-captured by a smaller, domain-focused corpus.

\textbf{Embedding structure and error analysis.}
The t-SNE projection in Figure~\ref{fig:tsne} confirms that LilyBERT's representations organise along musically meaningful dimensions: style clusters (Baroque, Classical, Romantic) form distinct regions, while within each cluster, individual composers occupy localised sub-regions. The confusion matrix in Figure~\ref{fig:confusion} reveals that most misclassifications occur between stylistically proximate composers --- for example, Haydn is occasionally confused with Mozart, and Grieg with Chopin --- suggesting that the model's errors reflect genuine musical similarity rather than random noise.

\textbf{Layer-wise analysis.}
Following prior work on BERT probing~\cite{tenney2019bert}, Figure~\ref{fig:layer_accuracy} reveals distinct layer dynamics for the two tasks. Composer classification peaks at the middle layers (6--9), consistent with the view that composer identity requires integrating both syntactic patterns (note sequences, voicing choices) and higher-level structural preferences. Style classification, by contrast, is best captured at early-to-middle layers (3--6) and degrades towards the output layer across all models. This suggests that stylistic features --- tempo conventions, instrumentation patterns, key preferences --- are encoded as relatively surface-level distributional regularities, while the deeper layers progressively specialise for the MLM objective at the expense of such global properties.

\section{Conclusion}\label{sec:conclusion}

We introduced BMdataset, a musicologically curated LilyPond dataset, and LilyBERT, a CodeBERT-based encoder adapted to symbolic music through vocabulary extension and MLM pre-training.

The central finding from our probing experiments is that 90M tokens of expert-curated data outperform 15B tokens of automatically converted data for both composer and style classification on the out-of-domain Mutopia corpus. Combining broad pre-training with domain-specific fine-tuning yields the best results overall (84.3\% composer accuracy), confirming that the two data regimes are complementary rather than redundant.

The dataset's principal limitation is its distributional skew towards Vivaldi, string instruments, and Late Baroque works. A further caveat is that the PDMX-LilyPond corpus used for continuous pre-training is produced by automatic conversion via \texttt{musicxml2ly}, which may introduce systematic artifacts (e.g., redundant overrides, non-idiomatic voicing) absent from hand-authored LilyPond; no post-conversion normalization was applied, and quantifying the impact of such artifacts remains an open question. Future work should expand the corpus to include additional periods and sources (e.g., Mutopia, IMSLP transcriptions), investigate structure-aware encoders that exploit LilyPond's parsable grammar (e.g., AST-based or graph-augmented models), apply LilyBERT embeddings to generative tasks such as LoRA-based LilyPond generation, and explore whether the hierarchical metadata can serve as conditioning signals for controllable music generation.

Because LilyPond is both a music engraving system and a structured programming language, its representations transfer to practical tasks such as score linting, style-aware auto-completion, and mixed-initiative editing. We release BMdataset, the extended tokenizer, and LilyBERT as a baseline for representation learning on LilyPond.

\textbf{Data availability.} \label{sec:data_availability}

All resources are publicly available to support reproducibility:
\begin{itemize}
    \item \textbf{Code}: \url{https://github.com/CSCPadova/lilybert}
    \item \textbf{Model weights}: \url{https://huggingface.co/csc-unipd/lilybert}
    \item \textbf{Dataset}: \url{https://doi.org/10.5281/zenodo.18723290}
\end{itemize}

	\begin{acknowledgments}
	We thank Mario Bolognani and the BaroqueMusic.it community for granting access to their LilyPond transcriptions. This work was partially funded by the European Union - NextGenerationEU, under the National Recovery and Resilience Plan (PNRR).
	\end{acknowledgments}

\bibliography{references}

\end{document}